\documentclass[12pt Times,a4paper]{article}
\usepackage[dvips]{graphicx}
\usepackage{multicol}
\newcommand{\bc}{\begin{center}}
\newcommand{\ec}{\end{center}}
\newcommand{\be}{\begin{equation}}
\newcommand{\ee}{\end{equation}} 
\newcommand{\ba}{\begin{eqnarray}}
\newcommand{\ea}{\end{eqnarray}}
\begin{document}
 \begin{flushright} INFN-GEF-TH-3/2013 \end{flushright}
 \vskip20pt
 \baselineskip 24pt
 \bc {\Large \bf CALCULATING THE DISSONANCE OF A CHORD ACCORDING TO HELMHOLTZ THEORY}
 \vskip20pt
\baselineskip 16pt 
\textbf{Giorgio Dillon }\\ 
 Dipartimento di Fisica, Universit\`a di Genova\\
INFN, Sezione di Genova \\
{\tt dillon@ge.infn.it}\ec
\vskip40pt
\begin{multicols}{2}
\baselineskip 14pt 
 \bc\textbf{ABSTRACT}\ec Following the ideas of von Helmholtz and Plomp-Levelt, an algorithm for calculating the   
dissonance  of complex sounds, free from logical inconsistencies and useful for comparing different chords, is 
proposed.  The method is tested by comparing different tunings of the same major triad.  Some interesting conclusions 
from this calculation may be drawn. \\ PACS: 43.75.Cd, 43.75.Bc
 \vskip30pt
 
 \bc\textbf{1. INTRODUCTION}\ec
According to von Helmholtz \cite{H} the dissonance between two simultaneously played complex tones is related to the 
roughness or rapid beating between their adjacent partials.
Because a musical sound is  harmonic (i.e. the frequencies of its partials are multiple of an audible fundamental 
frequency) it follows that the number of beating partials between two simultaneous musical sounds is minimized when 
the fundamental frequencies  are related by a ratio of small integers. This explains, on physical  grounds, the 
longstanding puzzle of why some musical intervals (those with simple ratios for their frequencies) are sensed as 
pleasant while others are judged  dissonant.

 After the precise definition of the  critical band provided by the work of Fletcher \cite{F}, Plomp and Levelt 
\cite{PL} reviewed and clarified the above explanation of consonance relating it to the \emph{critical bandwidth} 
(CB). They established experimentally a standard curve which represents the ``tonal consonance" between two pure 
(sinusoidal) tones, on an arbitrary scale, as a function of the ratio between the frequency difference and the CB. 
The maximal tonal dissonance was found at intervals of about 25\% of the CB at the mean frequency of the two tones. 
In Fig. 1 the function  $g(x)$ reproduces the Plomp-Levelt standard curve  
turned upside down to give dissonance instead of consonance. Accordingly the dissonance of a dyad of pure tones of 
frequencies $f_1$ and $f_2$ lying in the same CB $b(\bar{f})$ centered at the mean frequency, is  given by  
\be d(1,2)=g(x)\ee where $x=|f_1-f_2|/b(\bar{f})$.

\begin{figure}
   	\begin {center}\includegraphics[width=7cm]{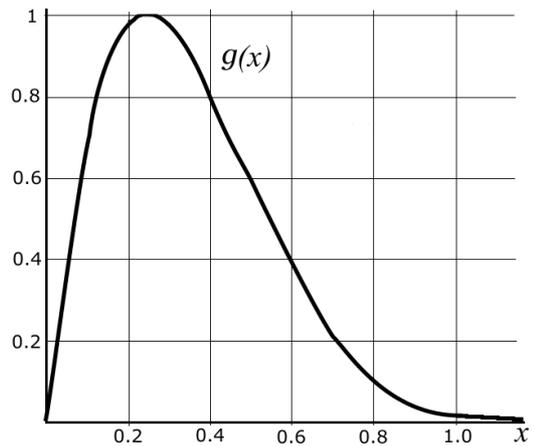}
   \end{center}
  \caption{The function that quantifies the dissonance experienced when two pure tones of frequencies $f_1$ and $f_2$ 
are sounded simultaneously, as derived by Plomp-Levelt (compare with Fig. 10 of \cite{PL}). The variable $x$ is 
defined as $x=|f_1-f_2|/b(\bar{f})$ where $b(\bar{f})$ is the CB at the mean frequency.}

   \label{1}
   \end{figure}

Given the dissonance of a single dyad, the problem arises how to calculate the total dissonance of a complex sound or 
of a chord of musical tones. The simplest assumption is the naive addition of the dissonances of each dyad present in 
the complex sound \cite{PL}. Anyhow, in summing the contributions  from the beating of  all nearby partials an 
appropriate weighting factor related to their intensities $I$, or better to their loudness $l$, is required:
\be d_{tot}=\sum_{i<j}f(l_i,l_j)d(i,j)
\label{2}\ee
 Different criteria have been adopted in the past. For instance, in  \cite{HK,Sj} each pure tone is supposed to 
contribute in proportion to its amplitude. In \cite{S} the weighting  factor is taken to be proportional to the 
loudness of the weaker component. 
\end{multicols} \twocolumn\baselineskip 14pt 
However we note that (\ref{2}) cannot be correct in general and that the above criteria hide a logical inconsistency. 
This may be seen as follows: Consider the dissonance of a dyad of pure tones as given by (\ref{1}). As a matter of 
fact one of them (or both) may be considered as the result of two superimposed pure tones with the same frequency and 
halved intensities. 
Then we could use (\ref{2}) to calculate the total dissonance and should recover the same result as (\ref{1}). This 
is not so. In fact, to avoid such a paradox, the weighting factor in (\ref{2}) should be proportional to the 
intensities (i.e. to the power per unit area) of both components, rather than to the amplitudes.

In \cite{KK} the concept of {\it dissonance intensity} was introduced as a physical counterpart of the psychological 
dissonance sensation and a power low is suggested to hold between the two according to the general psychophysics law 
of sensation postulated by Stevens \cite{Ste}. This is just the same kind of law that relates, for example, the 
intensity (or the sound pressure level) to the loudness, that will be exploited in Sec.2.  In this hypothesis 
Eq.(\ref{2}) does not hold since the summation concerns the dissonance intensities and not the dissonances as 
perceived by our hearing apparatus. Once the power law is established, the rule for calculating the total dissonance 
ensues. 
However, in spite of the accurate experiments and analysis that include the effect of the noise \cite{KK,KK2}, the 
resulting exponents lead to a framework that is plagued with analogous inconsitencies as (\ref{2}). Indeed one may  
check  the method calculating the effect of  two identical superimposed dyads (resulting in a total of 4 partials and 
6 dyads) whose dissonance output should be the same as an equivalent single dyad of pure tones with doubled 
intensities. Once again the discrepancy is sizable.

Following these premises we propose here a practical and mathematically consistent method for calculating the 
dissonance of a complex sound. Our recipe is intended to provide a relative measure for the dissonance useful to 
compare different chords. We shall test our method by comparing different tunings for the central major triad 
corresponding to historical temperaments. 

 \vskip30pt
\bc \textbf{2. WHICH EXPONENT?}\ec
 We begin this section giving a parameterization of the Plomp-Levelt curve of Fig.1. Though many parameterizations 
already exist \cite{Sj,S}, for our purpose the following simple polynomial fit is most suitable:
\be g(x)=Nx(1.2-x)^4 \label{F}\ee
that yields, with $N=4.906$, $g(x)=1$ at $x=0.25$. The fit (\ref{F}) assumes that $g(x)$ vanishes  identically at 
$x\geq 1.2$.

To avoid mathematical inconsistencies the $f(l_i,l_j)$ must necessarily be symmetric in the two indices and vanish 
when the loudness of one component vanishes. Then the simplest possible  form is
\be f(l_i,l_j)\propto (l_i\cdot l_j)^\alpha \label{lf}\ee
with $\alpha>0$.
 
 The loudness are approximately related to the intensities by \cite{KF}
\be l_{i,j}\propto \Big(\frac{I_{i,j}}{I^*}\Big)^{1/3} \label{lI}\ee
where $I^*$ is a reference intensity (depending on the frequency).

Now when two or more tones are sounded simultaneously, the total loudness depends on whether they lie or not within 
the same CB. If they do, the intensities are to be summed and the total loudness is
\be l_{tot} \propto \Big(\frac{1}{I^*}\sum_i I_i\Big)^{1/3}  \label{lt}\ee
In the opposite case the loudness should add \cite{KF}. However, since our parameterization for $g(x)$  vanishes 
identically outside the CB, we are not to worry about tones differing in frequency more than the relevant CB.

Then we are led to write  
\be d_{tot}=\frac{1}{l^{*2\alpha}}\Big(\sum_{i<j}(l_i\cdot l_j)^3d(i,j)^{3/\alpha} \Big)^{\alpha/3}\label{3}\ee
where $l^*$ is an arbitrary reference loudness.  Equation (\ref{3}), because of (\ref{lI}), is free from any 
drawbacks of the kind outlined in Sec.1 and consistent with the hypothesis put forward in \cite{KK}.  

 The problem is now how to choose $\alpha$. Note that for $\alpha=3$ one recovers the simple sum (\ref{2}) for the 
elementary dissonances $d(i,j)$ with the weighting factors proportional to the intensities, as already observed.
We tested different values for $\alpha$ and concluded that the most sensible one is $\alpha \approx 0.5$. The choice 
$\alpha=1/2$ is also appealing because it amounts to set
\be f(l_i,l_j)\propto\sqrt{l_il_j} \label{ml}\ee
that corresponds to a mean loudness for the dyad $(i,j)$.
 
\bc \textbf{3. COMPARING DIFFERENT TEMPERAMENTS FOR THE MAJOR TRIAD}\ec
The major triad is considered as the most consonant chord build up from the simultaneous sounding of three tones  and 
is most frequently encountered in tonal music. The fundamental frequencies of the three tones are in the ratios 
4:5:6, which correspond to the musical intervals of a major third (5/4), a minor third (6/5), and a perfect fifth 
(3/2). Actually these ratios refer to the \emph{just intonation} for these intervals, since, in practical usage, they  
must be slightly adjusted (\emph{tempered}). Though nowadays the so-called equal temperament that stems from the 
division of the octave in 12 equal parts has been definitively adopted in western music, different temperaments have 
been employed in the past according to the requirements and the taste of the music at the time. In this section we 
test (\ref{3}) (with $\alpha=1/2$) by comparing different tunings for the central major triad $(C_4-E_4-G_4)$. The 
notation $C_4$ singles out the note $C$ of the $4^{th}$ octave that corresponds to the $C$ at the middle of a piano 
keyboard.
 
The spectra of the three tones are supposed to be the same as that of a sawtooth wave characteristic of a bowed 
string \cite{FR} (the amplitude of the $n$th harmonic being $\propto 1/n$). We take the same loudness $l^*$ for the 
fundamentals and limit the summation to 6 harmonics. So the expression for this calculation takes the form
\be d_{tot}=\Big(\sum_{i<j}\sum_{m,n=1}^6\frac{1}{n^2m^2}\ d(n_i,m_j)^{6} \Big)^{1/6}\label{4}\ee
where $n_i$ stands for the $n$th harmonic of tone $i$ ($i,j=1,2,3$ ; 1 standing for $C_4$, 2 for $E_4$ and 3 for 
$G_4$  and $m,n=1,\ldots,6 $).
The values of the CB are taken from \cite{Z}. At this stage we neglected the frequency dependence of $I^*$ (see 
(\ref{lI})), that may equi\-valently be taken into account by some appropriate correction of the amplitudes in 
(\ref{4}). The results are displayed in Table 1.
 \begin{table} 
 {\small  
 \bc
 \begin{tabular}{l c c c c}
 \hline\hline
 &$C_4-E_4-G_4$& fund.&h.h.&$d_{tot}$\\
 Tuning&(frequencies in $Hz$)&x1000&x1000&x1000\\ \hline 
J.I.& $260-325-390$&328&415&430\\  
M.T. &$260-325-388.8$& 340 & 412 & 431\\  
W. III &$260-326-388.8$& 346 & 417 & 437\\  
Eq. &$260-327.6-389.6$& 352& 424 & 444\\  
Pyt. &$260-329-390$& 367 & 434 & 457\\ \hline\hline
\end{tabular}
\ec
}
\caption{The dissonance of the central major triad in different tunings: Just Intonation (J.I.), Meantone (M.T.), 
Werkmeister III (W. III), 12-Tone Equal Temperament (Eq.), Pythagorean (Pyt.). In the $2$nd  ($3$rd) column the 
contributions to the dissonance from the fundamentals (from the higher harmonics) is displayed separately. Each tone 
consists of 6 harmonics with amplitudes $\propto 1/n$.}
\end{table}
The central $C$ is tuned at $260Hz$ and the following temperaments are compared: Just Intonation (J.I.), Meantone 
(M.T.), Werkmeister III (W. III), 12-Tone Equal Temperament (Eq.), Pythago\-rean (Pyt.). In the $2$nd and $3$rd 
column the contributions to the dissonance from the fundamentals and from the higher harmonics is displayed 
separately. 
Of course no particular meaning should be attributed to the absolute value of the dissonances. What is meaningful 
here is that the calculation is accurate enough to discriminate among slightly different tunings of the same chord.

Some interesting features may be drawn from the above calculation. The first point is that, because of the large 
exponent in each term of the sum in (\ref{4}), only few terms are really important.  Specifically, in our case, the 
dyads leading the job are the following three:  $(n_i=4_1, n_j=3_2)$ ; $(n_i=1_2, n_j=1_3)$ ; $(n_i=5_2, n_j=4_3)$. 
The second point may be a surprise: In fact, the consonance of musical intervals with simple frequency ratios is 
usually ascribed to some coincidences in the spectra of the two tones. In the case of the major third the frequency 
ratio is 5/4 that means that the 5th harmonic of tone 1 coincides with the 4th harmonic of tone 2. Then the 
unpleasant effect of the Pythagorean third may be attributed to the beating of such harmonics not more in tune. Our 
calculation does not support this hypothesis. Instead it suggests that the Pythagorean major third is not as nice as 
that of just intonation mainly because the 3rd harmonic of tone 2 dangerously approaches the 4th harmonic of tone 1 
(rather than a missed coincidence of the 4th and the 5th harmonics). Of course things may drastically change with 
different spectra \cite{KK,S}.

\vskip30pt
\bc \textbf{4. CONCLUSIONS}\ec
In the present paper we aimed at giving a consistent algorithm for calculating and comparing the dissonances of 
composite sounds. Following the ideas of Plomp-Levelt \cite{PL} we proposed a recipe of how to ``sum" dissonances. 
Since physically one should sum the intensities, to avoid logical inconsistencies dissonances must be composed 
according to (\ref{3}). We chose $\alpha=1/2$. Actually the results are quite sensitive to this exponent. For 
instance for $\alpha=1$ the harmonics but the fundamentals would give a negligible contribution to the total 
dissonance (in contrast with von Helmholtz's hypothesis). Conversely too much weight would be attributed to higher 
partials for a lesser $\alpha$.  So $\alpha=1/2$ does seem a well gauged value. We think that this method may be 
useful for comparing chords from sounds with known spectra and for further analyses.



\vskip50pt
 
\begin{thebibliography}{99}
\small
\baselineskip12pt
\bibitem{H} H. von Helmholtz, {\it On the Sensation of Tone} English Ed., Dover, New York, 1954.  
 \bibitem{F} H. Fletcher, ``Auditory Patterns", {\it Rev. Mod. Phys.}, vol. 12, pp. 47-65, 1940.
 \bibitem{PL} R. Plomp and W.J.M. Levelt, ``Tonal Consonance and Critical Bandwidth", {\it J. Acoust. Soc. Am.}, vol. 
38, pp. 548-560, 1965.
 \bibitem{HK} W. Hutchinson and L. Knopoff, ``The acoustic measurement of consonance and dissonance for common 
practise triads", \emph{J. Musicological Research} Vol.3 6-22 (1979/80).
\bibitem{Sj} W.A. Sethares, ``Local consonance and the relationship between timbre and scales", \emph{J. Acoust. Soc. 
Am. } Vol.94 1218-1228 (1993).
\bibitem{S} W.A. Sethares, {\it Tuning, Timbre, Spectrum, Scale}, 2nd ed., Springer-Verlag, London 2005,  Chap. 6, p. 
99 and Appendices E-F.
 \bibitem{KK}A. Kameoka and M. Kurikawa, ``Consonance Theory Part II: Consonance of Complex Tones and Its Calculation 
Method", {\it J. Acoust. Soc. Am.}, vol.45, pp. 1460-1469, 1969.
 \bibitem{Ste} S.S. Stevens, ``On the Psychological Law" {\it Psychol. Rev.}, vol. 64, pp. 153-181, 1957.
 \bibitem{KK2}A. Kameoka and M. Kurikawa, ``Consonance Theory Part I: Consonance of Dyads", \emph{J. Acoust. Soc. 
Am.}, vol.45, pp. 1451-1459, 1969.
 \bibitem{KF} L.E. Kinsler, A.R. Frey, A.B. Coppens, J.V. Sanders, {\it Fundamentals of Acoustics}, 4th ed., John 
Wiley \& Sons, New York 2000, pp. 324-326.
\bibitem{FR} N.H. Fletcher and T.D. Rossing, \emph{The Physics of Musical Instruments}, 2nd ed., Springer, New York 
1998, Chap.10, p. 307.
\bibitem{Z} E. Zwicker, ``Subdivision of the Audible Frequency Range into Critical Bands", {\it J. Acoust. Soc. Am.}, 
vol. 33, p. 248, 1961.
 \end{thebibliography}
\end{document}